\documentclass[11pt]{article} 
\usepackage{hyperref}
\usepackage{graphicx} 
\pdfoutput=1 
\begin{document} 
\title{Convection rolls in a rotating box filled with beads} 
 
\author{Frank Rietz and Ralf Stannarius \\ 
\\\vspace{6pt} Dept. of Nonlinear Phenomena
\\ University of Magdeburg, Germany} 
 
\maketitle 
fluid dynamics video
\begin{abstract} 
Fill a box partially with a dry mixture of beads and rotate it slowly. As it is known from granular physics segregation patterns appear. Now put in more beads so that the box is almost completely filled and the beads can't move freely. Do we expect something unusual? Basically, only the quantity of beads was changed in the experiment but the results of that modification are clearly visible. Convection rolls are now observed that possess the same phenomenology as rolls in liquids or shaken granulates. The answer of that could not be easily defined and explained as in the cases of instable liquids that have a gradient (i.e. in density or temperature) or shaken systems that are strongly agitated. Besides the simplicity of the experiment known mechanisms for granular convection could not be applied. \qquad
\textbf{\href{http://ecommons.library.cornell.edu/bitstream/1813/14105/2/Rietz_high_quality_mpeg2.mpg}{Video}    
\qquad
\href{http://iep463.nat.uni-magdeburg.de/w3fr/PRL100_078002.pdf}{Paper}}
\begin{figure}[htbp]
	\centering
		\includegraphics[width=.85\columnwidth]{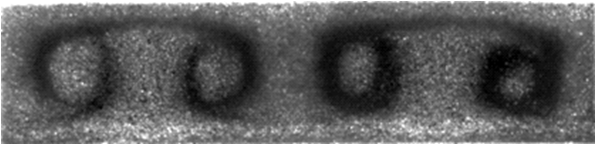}
	\label{fig:4rolls}
	\begin{quote}
	\caption{Example of 4 rolls in a rotating box that is only filled with a mixture of large and small beads. Dark/light areas indicate enrichment of small/large beads.}
	\end{quote}
\end{figure}

\end{abstract}

\end{document}